\documentclass[prl,superscriptaddress,twocolumn,nofootinbib,showpacs,preprintnumbers,floatfix]{revtex4}

\usepackage{mathrsfs}
\usepackage{amsfonts,amssymb,amsmath}
\usepackage{graphicx,epsfig}
\usepackage{multirow}
\usepackage{verbatim}

\usepackage{slashed}

\declareslashed{}{/}{0}{-0.05}{E}
\declareslashed{}{/}{0}{-0.05}{P}
\declareslashed{}{/}{0}{-0.10}{\Widehat{P}}

\def\fsu5{$\cal{F}$-$SU(5)$}

\def\bfsu5{$\boldsymbol{\mathcal{F}}$-$\boldsymbol{SU(5)}$}

\def\m1half{$M_{1/2}$}
\def\m3half{$M_{3/2}$}
\def\m32{$M_{32}$}

\def\mt2{$M_{T2}$}
\def\x2{$\chi^2$}
\def\2b{$M_{T2}b$}

\def\bs0{$B_S^0 \rightarrow \mu^+ \mu^-$}

\def\bea{\begin{eqnarray}}
\def\eea{\end{eqnarray}}

\def\nnb{\nonumber}

\begin{document}

\title{Naturalness in D-brane Inspired Models}

\author{Ron De Benedetti}

\affiliation{Department of Chemistry and Physics, Louisiana State University, Shreveport, Louisiana 71115 USA}

\author{Tianjun Li}

\affiliation{CAS Key Laboratory of Theoretical Physics, Institute of Theoretical Physics, 
Chinese Academy of Sciences, Beijing 100190, P. R. China}

\affiliation{ School of Physical Sciences, University of Chinese Academy of Sciences, 
No.19A Yuquan Road, Beijing 100049, P. R. China}

\author{James A. Maxin}

\affiliation{Department of Chemistry and Physics, Louisiana State University, Shreveport, Louisiana 71115 USA}

\author{Dimitri V. Nanopoulos}

\affiliation{George P. and Cynthia W. Mitchell Institute for Fundamental Physics and Astronomy, Texas A$\&$M University, College Station, TX 77843, USA}

\affiliation{Astroparticle Physics Group, Houston Advanced Research Center (HARC), Mitchell Campus, Woodlands, TX 77381, USA}

\affiliation{Academy of Athens, Division of Natural Sciences, 28 Panepistimiou Avenue, Athens 10679, Greece}


\begin{abstract}

We examine the naturalness of the D-brane inspired model constructed in flipped $SU(5)$ supplemented with vector-like particles at the TeV scale, dubbed flippons. We find the model can produce a mainly Higgsino-like lightest supersymmetric particle (LSP) and small light stops, as favored by naturalness. In fact, a large trilinear scalar $A_t$ term at the electroweak (EW) scale creates a large mass splitting between the top squarks, driving the light stop to near degeneracy with an LSP that is almost all Higgsino, with $\Delta M(\widetilde{t}_1, \widetilde{\chi}_1^0) < 5$~GeV, evading the LHC constraint on $\widetilde{t}_1 \to c \widetilde{\chi}_1^0$ thus far. Given the smallness of the light stop, generating a 125~GeV light Higgs boson mass is aided by one-loop contributions from the Yukawa couplings between the flippons and Higgs fields. The resulting parameter space satisfying naturalness is rather constrained, thus we assess its viability by means of comparison to the LHC constraint on soft charm jets and direction detection limits on spin-independent cross-sections. Finally, we compute the level of electroweak fine-tuning and uncover a region with $\Delta_{EW} < 30$, $i.e.$, fine-tuning better than 3\%, regarded as low electroweak fine-tuning. Given the small light stop, the electroweak fine-tuning from only the top squarks is of $\cal{O}$(1), indicating no fine-tuning from neither the light stop $\widetilde{t}_1$ nor the heavy stop $\widetilde{t}_2$.

\end{abstract}



\preprint{ACT-02-19, MI-TH-1920}

\maketitle


\section{Introduction}

The null results at the 13~TeV LHC Run 2 (LHC2) regarding the search for supersymmetry (SUSY) have now extended through 2018, as recent results find only Standard Model (SM) background events for data collected from 2016-18, inclusive of 137~${\rm fb}^{-1}$~\cite{Moriond_2019}. The rather strong limits derived from these observations though rely upon gluino ($\widetilde{g}$) and light stop ($\widetilde{t}_1$) channels producing hard jets via $\widetilde{g} \to \widetilde{t}_1 t \to t \bar{t} \widetilde{\chi}_1^0$ and $\widetilde{g} \to  q \bar{q} \widetilde{\chi}_1^0$, leading to limits on the gluino mass in excess of 2~TeV and on the light stop mass above 1~TeV. Such hard jets are easily accessible at the LHC2, affirming the rapid uninterrupted march to multi-TeV exclusion limits, yet approaching tension with SUSY's solution to the hierarchy problem, a prime reason motivating SUSY in the first place. On the other hand, the empty SUSY cupboard thus far prompts one to ask as to whether SUSY could be hiding in plain sight?

Natural SUSY, referred to as naturalness, strives for negligible electroweak fine-tuning in SUSY grand unified theory (GUT) models, defined by only natural cancellations amid terms in the tree-level minimization condition on the Higgs potential, plus radiative corrections. Insignificant amounts of fine-tuning require small terms in the minimization condition such that all terms on both sides of the equation are of comparable scale in order to compute the measured $Z$-boson mass. Such natural dynamics are very desirable, although an additional benefit can be realized that relates to the dilemma posed in the prior paragraph. The principal contributor to loop-level corrections are top squarks, so naturalness stresses small values for $M(\widetilde{t}_1)$, but small light stops could provoke degeneracy in the form of $M(\widetilde{t}_1) \sim M(\widetilde{\chi}_1^0)$. This raises an element of uncertainty for accessibility at the LHC, given the softness of these light stop events and hence unreliable distinction from the ubiquitous SM background. Indeed, one could expect these soft interactions to evade observation at the LHC if the light stop becomes rather compressed with the LSP. Furthermore, a complication surfaces with insufficient 1-loop and 2-loop SUSY contributions to the light Higgs boson mass $m_h$ from a small light stop, failing to generate the observed $m_h=125.09\pm 0.24$~GeV~\cite{Aad:2012tfa, Chatrchyan:2012xdj} light Higgs boson mass. We now introduce a model that has minimal electroweak fine-tuning but can handily achieve consistency with the light Higgs boson mass constraints, as well as other key experimental measurements, and could be flying just under the SUSY radar.

We shall study the flipped $SU(5)$ model~\cite{Barr:1981qv,Derendinger:1983aj,Antoniadis:1987dx} in this paper, where the gauge group $SU(5)\times U(1)_{X}$ can be embedded into the $SO(10)$ model. The flipped $SU(5)$ models can be constructed from the four-dimensional free fermionic string construction~\cite{Antoniadis:1988tt, Antoniadis:1989zy, Lopez:1992kg}, orbifold~\cite{Kim:2006hw, Huh:2009nh} and Calabi-Yau~\cite{Blumenhagen:2006ux} compactifications of the heterotic $E_8\times E_8$ string theory, intersecting D-brane model building~\cite{Chen:2005aba, Chen:2005cf, Chen:2006ip}, as well as F-theory model building~\cite{Jiang:2008yf, Jiang:2009za}. In addition, two of us (TL and DVN) with Jiang proposed the testable flipped $SU(5)$ models where the TeV-scale vector-like multiplets, dubbed flippons, are introduced, and string-scale gauge coupling unification can be achieved~\cite{Jiang:2006hf}. Such kind of models can be constructed from the four-dimensional free fermionic string construction~\cite{Lopez:1992kg}, intersecting D-brane model building\cite{Chen:2006ip}, as well as F-theory model building~\cite{Jiang:2008yf, Jiang:2009za}, and was referred to as \fsu5. This model persists in two classes: (i) the minimalistic formalism of the one-parameter version implementing vanishing No-Scale SUGRA soft SUSY breaking terms at the unification scale (For example, see Refs.~\cite{Li:2010ws,Maxin:2011hy,Li:2011ab,Li:2016bww} and references therein), and (ii) the general formalism with non-universal SUSY soft breaking terms mirroring the flipped $SU(5)$ GUT representation, inspired by D-brane model building, and thus informally designated the \fsu5 D-brane inspired model~\cite{DeBenedetti:2018fxa}. This second approach endures as merely a D-brane $inspired$ model and not a formally constructed D-brane model by reason of forbidden Yukawa coupling terms in the Higgs and Yukawa superpotentials, though we discuss in the next section possible methods to elude these hurdles. The \fsu5 D-brane inspired model revealed a possible region of naturalness featuring small light stops and a Higgsino-like lightest supersymmetric particle (LSP)~\cite{DeBenedetti:2018fxa}, which we shall more fully unpack here in this work. For a discussion of naturalness in a Pati-Salam model constructed from intersecting D6-branes in Type IIA string theory, see Ref.~\cite{Ahmed:2017ttx}.

Fine-tuning in the minimalistic formalism of \fsu5 has been explored~\cite{Leggett:2014hha}. In Ref.~\cite{Leggett:2014hha} it was shown that the contemporary measures of fine-tuning we shall employ in this analysis are essentially structurally similar to an original fine-tuning measure, $\Delta_{\rm EENZ}$~\cite{Ellis:1986yg, Barbieri:1987fn}, first prescribed some 30 years ago by Ellis, Enqvist, Nanopoulos, and Zwirner (EENZ). The one-parameter version of the model possesses an intrinsic proportional dependence of all model scales on the unified gaugino mass parameter $M_{1/2}$, inclusive of the $Z$-boson mass expressed as a simple quadratic function of $M_{1/2}$. The implication was electroweak fine-tuning of unity scale~\cite{Leggett:2014hha}. The minimalistic version of \fsu5 is presently under probe at the LHC2~\cite{Li:2017kcq} and has thus far survived the 13~TeV LHC2 137~${\rm fb}^{-1}$ results~\cite{Moriond_2019}. Now we turn our attention to the less internally constrained version of \fsu5, evaluating fine-tuning in the D-brane inspired model. Our goal here is to show that this class of \fsu5 is inflicted with a minimal amount of fine-tuning also, and even though the one-parameter version is presently experiencing a direct probe by the LHC, the naturalness sector of the D-brane inspired model has been just $under$ the reach of the LHC2. 

In this work we first supply a brief review of the flipped $SU(5)$ class of models and the D-brane inspired model in particular. Then we delve into the comprehensive numerical procedure necessary to investigate naturalness. Once the numerical approach has been wholly dissected, we expand upon the phenomenology of the naturalness sector and the attainment of small light stops, Higgsino-like LSPs, and other associated provisions essential for low fine-tuning, accompanied by light Higgs boson masses lifted to 125~GeV for many points by the vector-like flippon contributions. Integrated into this analysis will be evidence of our naturalness sector skirting under the LHC constraints up to this point, and moreover, an evaluation against dark matter direct detection experiments and application of their results as a constraint on the naturalness region. Finally, we conclude with the fine-tuning calculations and assessment of the numerical findings.

\section{Review of \bfsu5 Model}

We review here only the primary principles of \fsu5. In the minimal flipped $SU(5)$ model~\cite{Barr:1981qv,Derendinger:1983aj,Antoniadis:1987dx}, the gauge group $SU(5)\times U(1)_{X}$ can be embedded within the $SO(10)$ model. Please see Refs.~\cite{Maxin:2011hy,Li:2011ab,Li:2013naa,Leggett:2014hha,Li:2016bww} and references therein for a more in-depth analysis of the minimal flipped $SU(5)$ model. The generator $U(1)_{Y'}$ in $SU(5)$ is defined as 
\bea 
T_{\rm U(1)_{Y'}}={\rm diag} \left(-\frac{1}{3}, -\frac{1}{3}, -\frac{1}{3},
 \frac{1}{2},  \frac{1}{2} \right)~,~\,
\label{u1yp}
\eea
and as a result the hypercharge is given by
\bea
Q_{Y} = \frac{1}{5} \left( Q_{X}-Q_{Y'} \right).
\label{ycharge}
\eea
There are three families of the SM fermions with quantum numbers under $SU(5)\times U(1)_{X}$ given by, respectively,
\bea
F_i={\mathbf{(10, 1)}},~ {\bar f}_i={\mathbf{(\bar 5, -3)}},~
{\bar l}_i={\mathbf{(1, 5)}},
\label{smfermions}
\eea
where $i=1, 2, 3$. The SM particle assignments in $F_i$, ${\bar f}_i$ and ${\bar l}_i$ are
\bea
F_i=(Q_i, D^c_i, N^c_i),~{\overline f}_i=(U^c_i, L_i),~{\overline l}_i=E^c_i~,~
\label{smparticles}
\eea
where $Q_i$, $U^c_i$, $D^c_i$,  $L_i$, $E^c_i$ and $N^c_i$ are the left-handed quark doublets, right-handed up-type quarks, down-type quarks, left-handed lepton doublets, right-handed charged leptons, and neutrinos, respectively. The introduction of three SM singlets $\phi_i$ can generate the heavy right-handed neutrino masses.

The GUT and electroweak gauge symmetries are broken through the introduction of the two pairs of Higgs representations
\begin{eqnarray}
H&=&{\mathbf{(10, 1)}},~{\overline{H}}={\mathbf{({\overline{10}}, -1)}}, \nonumber \\
h&=&{\mathbf{(5, -2)}},~{\overline h}={\mathbf{({\bar {5}}, 2)}}.
\label{Higgse1}
\end{eqnarray}
The $H$ multiplet states are labeled by the same symbols as the $F$ multiplet, and for ${\overline H}$ we only add a ``bar'' above the fields. Specifically, the Higgs particles are
\bea
H=(Q_H, D_H^c, N_H^c)~,~
{\overline{H}}= ({\overline{Q}}_{\overline{H}}, {\overline{D}}^c_{\overline{H}}, 
{\overline {N}}^c_{\overline H})~,~\,
\label{Higgse2}
\eea
\bea
h=(D_h, D_h, D_h, H_d)~,~
{\overline h}=({\overline {D}}_{\overline h}, {\overline {D}}_{\overline h},
{\overline {D}}_{\overline h}, H_u)~,~\,
\label{Higgse3}
\eea
where $H_d$ and $H_u$ are one pair of Higgs doublets in the MSSM.

The ensuing Higgs superpotential at the GUT scale breaks the $SU(5)\times U(1)_{X}$ gauge symmetry down to the SM gauge symmetry 
\bea
{\it W}_{\rm GUT}=\lambda_1 H H h + \lambda_2 {\overline H} {\overline H} {\overline
h} + \Phi ({\overline H} H-M_{\rm H}^2)~.~ 
\label{spgut} 
\eea
Merely one F-flat and D-flat direction exists, and that can be rotated along the $N^c_H$ and ${\overline {N}}^c_{\overline H}$ directions. Consequently, we have $<N^c_H>=<{\overline {N}}^c_{\overline H}>=M_{\rm H}$. With the exception of $D_H^c$ and ${\overline {D}}^c_{\overline H}$, the superfields $H$ and ${\overline H}$ are ``eaten'' and acquire substantial masses via the supersymmetric Higgs mechanism. Moreover, the superpotential terms $ \lambda_1 H H h$ and $ \lambda_2 {\overline H} {\overline H} {\overline h}$ couple $D_H^c$ and ${\overline {D}}^c_{\overline H}$ respectively with $D_h$ and ${\overline {D}}_{\overline h}$, which forms massive eigenstates with masses $2 \lambda_1 <N_H^c>$ and $2 \lambda_2 <{\overline {N}}^c_{\overline H}>$. Therefore, the doublet-triplet splitting due to the missing partner mechanism~\cite{Antoniadis:1987dx} naturally arises. However, the triplets in $h$ and ${\overline h}$ only have a small mixing via the $\mu$ term, so the colored Higgsino-exchange mediated proton decay is negligible, {\it i.e.}, there is no dimension-5 proton decay problem. 

The following vector-like particles (flippons) at the TeV scale are introduced to realize string-scale gauge coupling unification~\cite{Jiang:2006hf, Jiang:2008yf, Jiang:2009za} 
\begin{eqnarray}
&& XF ={\mathbf{(10, 1)}}~,~{\overline{XF}}={\mathbf{({\overline{10}}, -1)}}~,~\nnb \\
&& Xl={\mathbf{(1, -5)}}~,~{\overline{Xl}}={\mathbf{(1, 5)}}~.~\,
\end{eqnarray}
The particle content from the decompositions of $XF$, ${\overline{XF}}$, $Xl$, and ${\overline{Xl}}$ under the SM gauge symmetry are
\begin{eqnarray}
&& XF = (XQ, XD^c, XN^c)~,~ {\overline{XF}}=(XQ^c, XD, XN)~,~\nnb \\
&& Xl= XE~,~ {\overline{Xl}}= XE^c~.~
\end{eqnarray}
The quantum numbers for the extra vector-like particles under the $SU(3)_C \times SU(2)_L \times U(1)_Y$ gauge symmetry are
\begin{eqnarray}
&& XQ={\mathbf{(3, 2, \frac{1}{6})}}~,~
XQ^c={\mathbf{({\bar 3}, 2,-\frac{1}{6})}} ~,~\\
&& XD={\mathbf{({3},1, -\frac{1}{3})}}~,~
XD^c={\mathbf{({\bar 3},  1, \frac{1}{3})}}~,~\\
&& XN={\mathbf{({1},  1, {0})}}~,~
XN^c={\mathbf{({1},  1, {0})}} ~,~\\
&& XE={\mathbf{({1},  1, {-1})}}~,~
XE^c={\mathbf{({1},  1, {1})}}~.~\,
\label{qnum}
\end{eqnarray}
The superpotential is
\bea 
{ W}_{\rm Yukawa} &=&  y_{ij}^{D}
F_i F_j h + y_{ij}^{U \nu} F_i  {\overline f}_j {\overline
h}+ y_{ij}^{E} {\overline l}_i  {\overline f}_j h  
\nnb \\ &&
+ \mu h {\overline h}
+ y_{ij}^{N} \phi_i {\overline H} F_j +M_{ij}^{\phi} \phi_i \phi_j
\nnb \\ &&
+ y_{XF} XF XF h + y_{\overline{XF}} {\overline{XF}} {\overline{XF}} {\overline h}
\nnb \\ &&
+ M_{XF} {\overline{XF}}  XF + M_{Xl} {\overline{Xl}}  Xl
~,~\,
\label{potgut}
\eea
and the above superpotential after the $SU(5)\times U(1)_X$ gauge symmetry is broken down to the SM gauge symmetry gives
\bea 
{ W_{SSM}}&=&
y_{ij}^{D} D^c_i Q_j H_d+ y_{ji}^{U \nu} U^c_i Q_j H_u
+ y_{ij}^{E} E^c_i L_j H_d \nnb \\ &&
+  y_{ij}^{U \nu} N^c_i L_j H_u  +  \mu H_d H_u+ y_{ij}^{N} 
\langle {\overline {N}}^c_{\overline H} \rangle \phi_i N^c_j
\nnb \\ &&
+ y_{XF} XQ XD^c H_d + y_{\overline{XF}} XQ^c XD H_u
\nnb \\ &&
+M_{XF}\left(XQ^c XQ + XD^c XD\right) 
\nnb \\ &&
+ M_{Xl} XE^c  XE+M_{ij}^{\phi} \phi_i \phi_j
\nnb \\ &&
 + \cdots (\textrm{decoupled below $M_{GUT}$}). 
\label{poten1}
\eea
where $y_{ij}^{D}$, $y_{ij}^{U \nu}$, $y_{ij}^{E}$, $y_{ij}^{N}$, $y_{XF} $, and $y_{\overline{XF}}$ are Yukawa couplings, $\mu$ is the bilinear Higgs mass term, and $M_{ij}^{\phi}$, $M_{XF} $ and $M_{Xl}$ are masses for new particles. The new particles are the vector-like flippons, though we shall not formally compute the masses $M_{ij}^{\phi}$, $M_{XF}$ ,and $M_{Xl}$ in this study, reserving this analysis for the future.  Only a common mass decoupling scale $M_V$ for the flippon vector-like particles is enforced. Present LHC constraints on vector-like $T$ and $B$ quarks~\cite{atlas-vectorlike} fix lower limits of about 855~GeV for $(XQ, ~XQ^c)$ vector-like flippons and 735~GeV for $(XD, ~XD^c)$ vector-like flippons. We therefore suitably place our lower $M_V$ limit at $M_V \ge 855$~GeV to guarantee inclusion of all experimentally viable flippon masses in our work.

The two-stage unification of flipped $SU(5)$~\cite{Barr:1981qv,Derendinger:1983aj,Antoniadis:1987dx} allows for fundamental GUT scale Higgs representations (not adjoints), natural doublet-triplet splitting, suppression of dimension-five proton decay~\cite{Harnik:2004yp}, and a two-step see-saw mechanism for neutrino masses~\cite{Ellis:1992nq,Ellis:1993ks}. More precisely, a distinct separation between the ultimate $SU(5) \times U(1)_X$ unification at around $3 \times 10^{17}$~GeV and the penultimate $SU(3)_C \times SU(2)_L$ unification near $10^{16}$~GeV emerges due to revisions to the one-loop gauge $\beta$-function coefficients $b_i$ to include contributions from the vector-like flippon multiplets that induce the required flattening of the $SU(3)$ Renormalization Group Equation (RGE) running ($b_3 = 0$)~\cite{Li:2010ws}. The $M2$ and $M3$ gaugino mass terms are unified into a single mass term $M5 = M2 = M3$~\cite{Li:2010rz}, and hence $\alpha_5 = \alpha_2 = \alpha_3$, at the $SU(3)_C \times SU(2)_L$ unification near $10^{16}$~GeV. The $M1$ gaugino mass term runs up to the $SU(5) \times U(1)_X$ unification at $M_{\cal F}$, by virtue of a small shift due to $U(1)_X$ flux effects~\cite{Li:2010rz} between the $SU(3)_C \times SU(2)_L$ unification around $10^{16}$~GeV and the $SU(5) \times U(1)_X$ unification around $3 \times10^{17}$~GeV~\cite{Li:2010ws}. This shift motivates that the $M1$ gaugino mass term above the unification around $10^{16}$~GeV be referred to as $M_{1X}$. The scale $M_{\cal F}$ is defined by unification of the couplings $\alpha_5 = \alpha_{1X}$, which boosts unification to near the string scale and Planck mass. The flattening of the $M3$ gaugino mass dynamic evolution down to the electroweak scale generates the mass texture of $M(\widetilde{t}_1) < M(\widetilde{g}) < M(\widetilde{q})$, with the light stop and gluino lighter than all other squarks~\cite{Li:2011ab}.

The SUSY breaking soft terms at the $M_{\cal F}$ scale in the ${\cal F}$-$SU(5)$ model are appropriately $M_5$, $M_{1X}$, $M_{U^c L}$, $M_{E^c}$, $M_{Q D^c N^c}$, $M_{H_u}$,  $M_{H_d}$, $A_{\tau}$, $A_t$, and $A_b$. Non-universal SUSY breaking soft terms such as these are inspired partially by D-brane model building~\cite{Chen:2006ip}, where $F_i$, ${\overline f}_i$, ${\overline l}_i$, and $h/{\overline h}$ result from intersections of different stacks of D-branes. In this event, the associated SUSY breaking soft mass terms and trilinear scalar $A$ terms are different, while $M_{H_u}$ is equal to $M_{H_d}$. Despite the fact the Yukawa terms $H H h$ and ${\overline H} {\overline H} {\overline h}$ of Eq. (\ref{spgut}) and $F_i F_j h$, $XF XF h$, and $\overline{XF} \overline{XF} {\overline h}$ of Eq.~(\ref{potgut}) are forbidden by the anomalous global $U(1)$ symmetry of $U(5)$, these Yukawa terms could be generated from high-dimensional operators or instanton effects. In fact, the $SU(5) \times U(1)_X$ models differ from $SU(5)$ models such that in \fsu5 the Yukawa term $F_i F_j h$ gives down-type quark masses, so their Yukawa couplings can be small and be generated via high-dimensional operators or instanton effects.

\begin{figure}[t]
       \centering
        \includegraphics[width=0.5\textwidth]{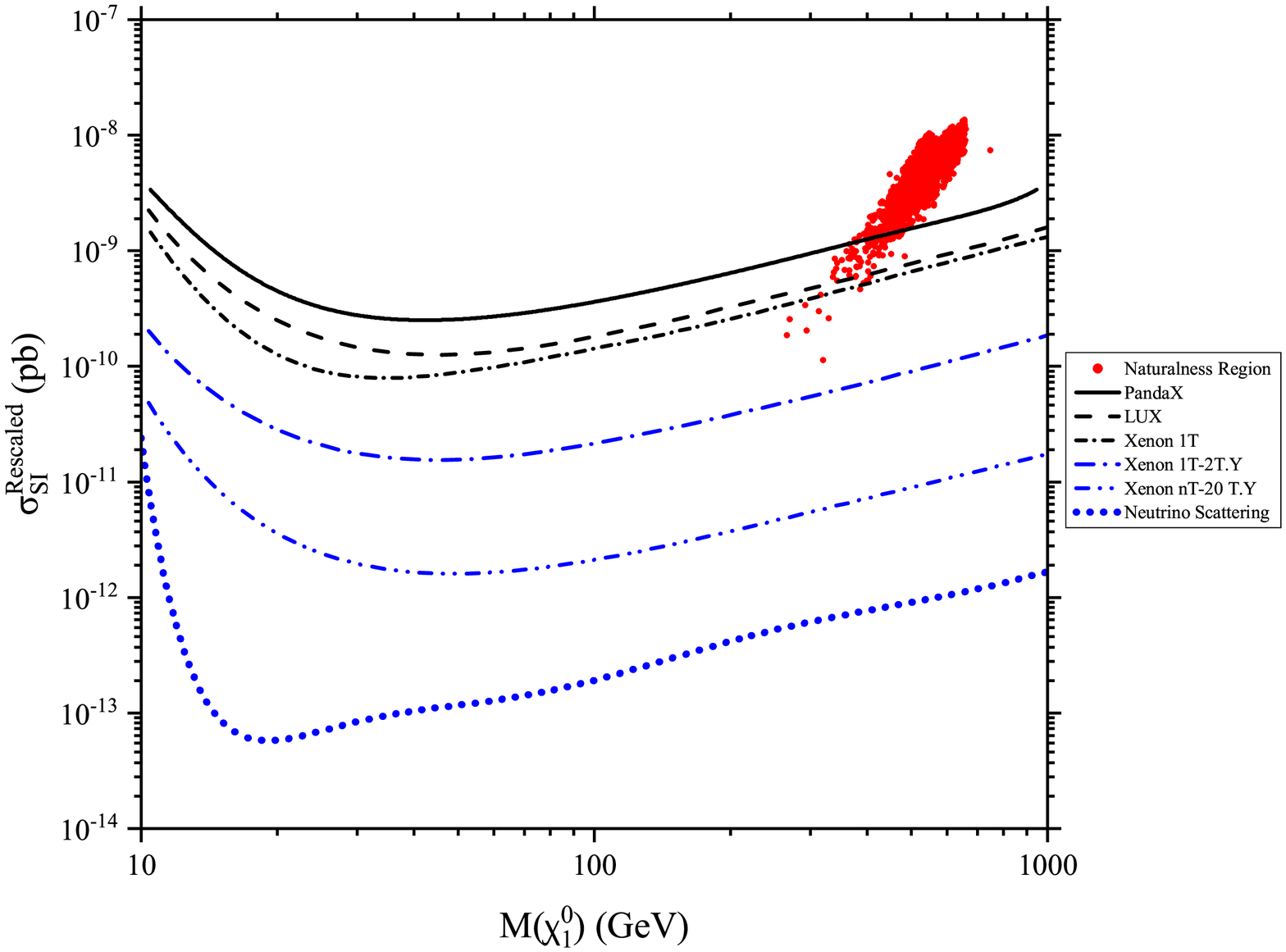}
        \caption{Depiction of the \fsu5 D-brane inspired naturalness region, represented by the small red points. The LUX, PandaX-II, and XENON100 dark matter direct detection upper limits on spin-independent neutralino-nucleus cross-sections are sketched and labeled, as well the neutrino scattering floor. The $\sim 2900$ points here satisfy the constraints $M({\widetilde{g}}) \ge 1.6$~TeV, $124 \le m_h \le 128$~GeV, $\Omega h^2 \le 0.1221$, $\Delta M(\widetilde{t}_1,\widetilde{\chi}_1^0) \le 10$~GeV, and $\widetilde{\chi}_1^0 > 80\%$ Higgsino. The cross-sections $\sigma_{SI}^{\rm Rescaled}$ are rescaled in accordance with Eq.~(\ref{eq:omega}).}
        \label{fig:naturalness}
\end{figure}

\begin{figure}[t]
       \centering
        \includegraphics[width=0.5\textwidth]{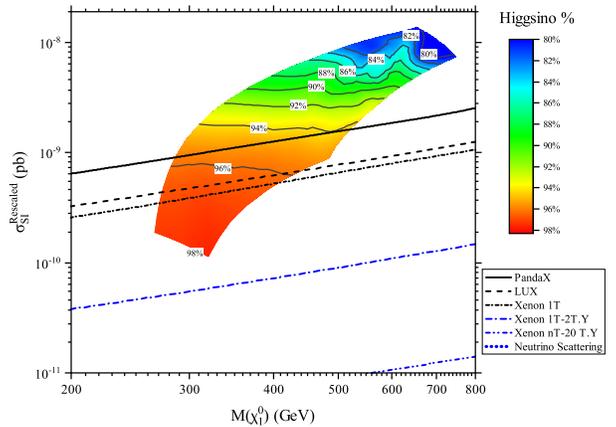}
        \caption{Smoothly flowing contours of the Higgsino-like LSP percentage within the \fsu5 D-brane inspired naturalness region. The LUX, PandaX-II, and XENON100 dark matter direct detection upper limits on spin-independent neutralino-nucleus cross-sections are sketched and labeled. The region illustrated here satisfies the constraints $M({\widetilde{g}}) \ge 1.6$~TeV, $124 \le m_h \le 128$~GeV, $\Omega h^2 \le 0.1221$, $\Delta M(\widetilde{t}_1,\widetilde{\chi}_1^0) \le 10$~GeV, and $\widetilde{\chi}_1^0 > 80\%$ Higgsino. The cross-sections $\sigma_{SI}^{\rm Rescaled}$ are rescaled in accordance with Eq.~(\ref{eq:omega}).}
        \label{fig:higgsino}
\end{figure}

\section{Numerical Approach}

\begin{figure}[t]
       \centering
        \includegraphics[width=0.5\textwidth]{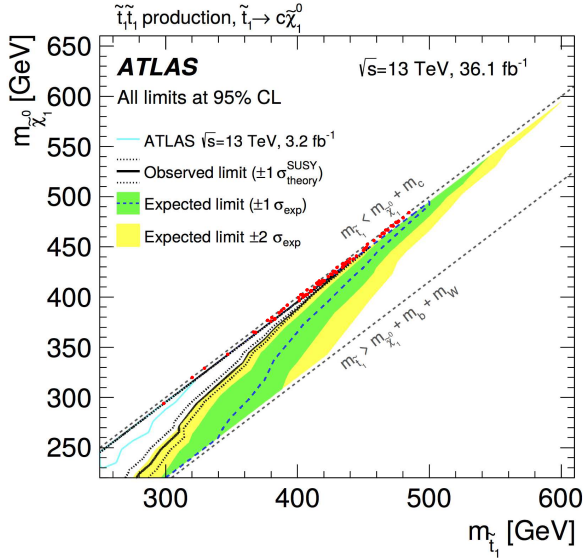}
        \caption{Layered illustration of the \fsu5 D-brane inspired naturalness region (represented by the small red points) superimposed upon the ATLAS Collaboration exclusion curve plot on $\widetilde{t}_1 \widetilde{t}_1$ production in the monojet search region for the channel $\widetilde{t}_1 \to c \widetilde{\chi}_1^0$, reprinted from Ref.~\cite{Aaboud:2017phn}. The branching fraction for $\widetilde{g} \to \widetilde{t}_1 t$ and $\widetilde{t}_1 \to c \widetilde{\chi}_1^0$ is nearly 100\% for our points shown in this Figure. The naturalness region in this diagram involves the 74 points that satisfy  $M({\widetilde{g}}) \ge 1.6$~TeV, $124 \le m_h \le 128$~GeV, $\Omega h^2 \le 0.1221$, $\sigma_{SI}^{\rm Rescaled} \le 1.5 \times 10^{-9}$~pb, and $\Delta M(\widetilde{t}_1,\widetilde{\chi}_1^0) \le 5$~GeV. All 74 points depicted here possess an LSP that is at least 92\% Higgsino, but no more than 98\% Higgsino.}
        \label{fig:atlas1}
\end{figure}

\begin{figure}[t]
       \centering
        \includegraphics[width=0.5\textwidth]{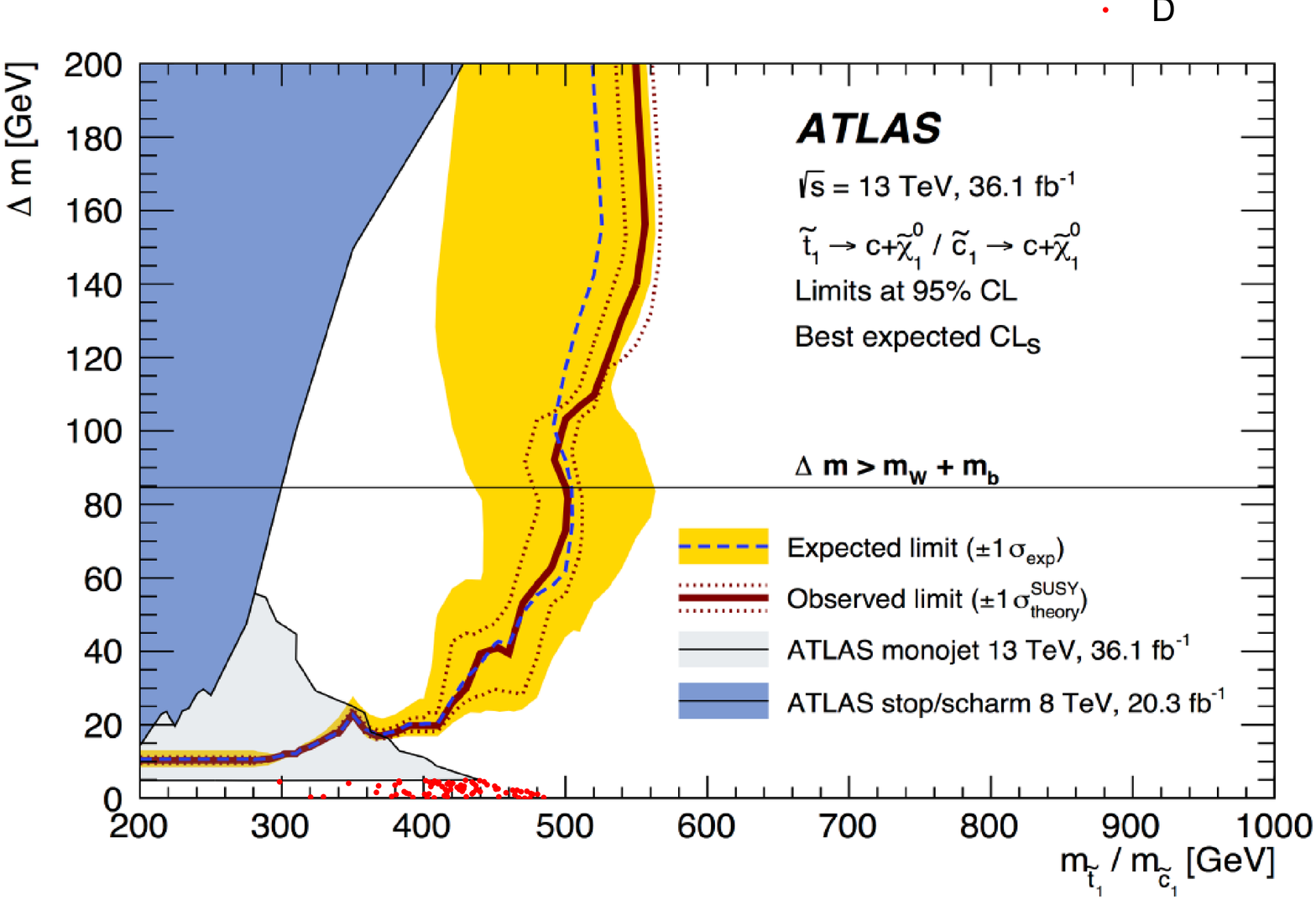}
        \caption{Layered illustration of the \fsu5 D-brane inspired naturalness region (represented by the small red points) superimposed upon the ATLAS exclusion curve plot in the charm jets plus zero lepton (0L) search region for the channel $\widetilde{t}_1 \to c \widetilde{\chi}_1^0$, reprinted from Ref.~\cite{Aaboud:2018zjf}. This ATLAS exclusion plot also includes the monojet search region of Ref.~\cite{Aaboud:2017phn}. The branching fraction for $\widetilde{g} \to \widetilde{t}_1 t$ and $\widetilde{t}_1 \to c \widetilde{\chi}_1^0$ is nearly 100\% for our points shown in this Figure. The naturalness region in this diagram involves the 74 points that satisfy  $M({\widetilde{g}}) \ge 1.6$~TeV, $124 \le m_h \le 128$~GeV, $\Omega h^2 \le 0.1221$, $\sigma_{SI}^{\rm Rescaled} \le 1.5 \times 10^{-9}$~pb, and $\Delta M(\widetilde{t}_1,\widetilde{\chi}_1^0) \le 5$~GeV. All 74 points depicted here possess an LSP that is at least 92\% Higgsino, but no more than 98\% Higgsino.}
        \label{fig:atlas2}
\end{figure}

At the unification scale of $M_{\cal F} \sim 3 \times 10^{17}$~GeV, the \fsu5 general SUSY breaking soft terms are applied, namely $M_5$, $M_{1X}$, $M_{U^c L}$, $M_{E^c}$, $M_{Q D^c N^c}$,  $M_{H_u} = M_{H_d}$, $A_{\tau}$, $A_t$, and $A_b$. The \fsu5 unification scale $M_{\cal F}$ near the string and Planck scale is in contrast to the usual lower GUT scale of about $10^{16}$~GeV in the MSSM. All SUSY breaking soft terms are allowed to float up to 5~TeV, with the $A$ terms varying between $\pm5$~TeV, though specifically for the $A_t$ term we establish an extended lower limit of -7~TeV. A $\pm1.5$~GeV margin of error is permitted around the top quark world average of 173.2~GeV~\cite{CDF:2013jga}. The ratio of the vacuum energy expectation values tan$\beta$ is allowed to span its entire range of $5 \le {\rm tan}\beta \le 60$. The flippon vector-like decoupling scale is sampled within the range $855 \le M_V \le 23,000$~GeV. We adopt $\mu >0$ for all points as suggested by the results of $g_{\mu}-2$ for the muon.

The model is constrained to be consistent with both the WMAP 9-year~\cite{Hinshaw:2012aka} and Planck 2018~\cite{Aghanim:2018eyx} relic density measurements, imposing an upper limit of $\Omega h^2 \le 0.1221$. Given the large annihilation cross-section of a Higgsino-like LSP, no lower limit is placed on $\Omega h^2$. The strongest LHC gluino limits arise from the search regions $\widetilde{g} \to \widetilde{q} q \to q \bar{q} \widetilde{\chi}_1^0$ and $\widetilde{g} \to \widetilde{t}_1 t \to t \bar{t} \widetilde{\chi}_1^0$, however, in our study here we are interested in the channel producing a top+charm via $\widetilde{g} \to \widetilde{t}_1 t \to c t \widetilde{\chi}_1^0$, which persists with weaker limits. Accordingly, we implement a somewhat weaker lower boundary of $M({\widetilde{g}}) \ge 1.6$~TeV given that these gluinos are not easily accessible.

The theoretical calculation of the light Higgs boson mass is allowed to vary from the experimental central value of $m_h = 125.09$~GeV, where we account for a 2$\sigma$ experimental uncertainty and theoretical uncertainty of 1.5 GeV. The allocated range for the flippon Yukawa coupling spans from its minimal value (no coupling between the flippons and Higgs fields) to its maximal value (maximum coupling between flippons and Higgs fields). In the maximum case, the light Higgs boson mass calculation consists of the 1-loop and 2-loop SUSY contributions, mainly from the coupling to the light stop, plus the vector-like flippon contributions. This maximal value implies the $(XD,~XD^c)$ Yukawa coupling is fixed at $Y_{XD} = 0$ and the $(XU,~XU^c)$ Yukawa coupling is set at $Y_{XU} = 1$, while the $(XD,~XD^c)$ trilinear coupling $A$ term set at $A_{XD} = 0$ and the $(XU,~XU^c)$ $A$ term is fixed at $A_{XU} = A_U = A_0$~\cite{Huo:2011zt,Li:2011ab}. In total, after including all contributions, the light Higgs boson mass calculation must return a value of $124 \le m_h \le 128$~GeV.

We further assess the model against rare decay processes, to include the branching ratio of the rare b-quark decay of $Br(b \to s \gamma) = (3.43 \pm 0.21^{stat}~ ±\pm 0.24^{th} \pm 0.07^{sys}) \times 10^{-4}$~\cite{HFAG}, the branching ratio of the rare B-meson decay to a dimuon of $Br(B_s^0 \to \mu^+ \mu^-) = (2.9 \pm 0.7 \pm 0.29^{th}) \times 10^{-9}$~\cite{CMS:2014xfa}, and the 3$\sigma$ intervals around the Standard Model result and experimental measurement of the SUSY contribution to the anomalous magnetic moment of the muon of $-17.7 \times10^{-10} \le \Delta a_{\mu} \le 43.8 \times 10^{-10}$~\cite{Aoyama:2012wk}. We only inspect the model versus these rare decay processes, and do not explicitly constrain the model per these experimental limits.

The naturalness region is also evaluated against dark matter direct detection constraints on spin-independent cross-sections $\sigma_{SI}$ for neutralino-nucleus interactions established by the Large Underground Xenon (LUX) experiment~\cite{Akerib:2016vxi}, PandaX-II Experiment~\cite{Tan:2016zwf}, and XENON100 Collaboration~\cite{Aprile:2018dbl}. The relic density calculations involve only the SUSY lightest neutralino $\widetilde{\chi}_1^0$ abundance, hence all points must admit alternative components to maintain compatibility with the WMAP 9-year and 2018 Planck total observed relic density, thus the spin-independent cross-section calculations are rescaled as follows:
\bea
\sigma^{\rm Rescaled}_{SI}=\sigma_{SI}\frac{\Omega h^2}{0.1200}~.~
\label{eq:omega}
\eea

The 150 million points scanned in Ref.~\cite{DeBenedetti:2018fxa} were enhanced in this effort by an additional 250 million points. The Higgs and SUSY mass spectra, relic density, dark matter direct detection cross-sections, LSP composition, and rare decay processes are calculated with {\tt MicrOMEGAs~2.1}~\cite{Belanger:2008sj} employing a proprietary mpi modification of the {\tt SuSpect~2.34}~\cite{Djouadi:2002ze} codebase to run flippon and general No-Scale ${\cal F}$-$SU(5)$ enhanced RGEs, implementing non-universal soft supersymmetry breaking parameters at the $M_{\cal F}$ scale. Supersymmetric particle decays are calculated with {\tt SUSY-HIT~1.5a}~\cite{Djouadi:2006bz}. The Particle Data Group~\cite{Tanabashi:2018oca} world average for the strong coupling constant is $\alpha_S (M_Z) = 0.1181 \pm 0.0011$ at 1$\sigma$, and we assume a value in this work of $\alpha_S = 0.1184$.

\section{Naturalness Phenomenology}

\begin{table*}[htp]
  \centering
  \scriptsize
  \caption{The SUSY breaking soft terms, in addition to the vector-like flippon decoupling scale $M_V$, the low energy ratio of Higgs vacuum expectation values (VEVs) tan$\beta$, and top quark mass $M_t$ for the  \fsu5 D-brane inspired naturalness region. Each benchmark point is identified with an alphabetical label in order to link the data in TABLE~\ref{tab:spectra1} with the data in TABLES~\ref{tab:spectra2} - \ref{tab:spectra3}. All masses are in GeV. The relic density $\Omega h^2$, rescaled dark matter direct detection cross section $\sigma_{SI}^{\rm Rescaled}$ (in pb), and Higgsino percentage of the LSP are also given. }
\label{tab:spectra1}
\begin{tabular}{|c||c|c|c|c|c|c|c|c|c|c|c|c||c|c|c|} \hline
$ {\rm Benchmark} $ & $M_5$ & $M_{1{\rm X}}$  &  $M_{U^c L}$  &  $M_{E^c}$  &  $M_{Q D^c N^c}$  &  $M_{H_u} = M_{H_d}$  &  $A_{\tau}$  &  $A_t$  &  $A_b$  &   $M_V$  &  $ {\rm tan}\beta $  &  $M_t$ & $\Omega h^2$ & $\sigma_{SI}^{\rm Rescaled}$ & ${\rm LSP}$ \\ \hline \hline
$	{\rm A }	$&$	1500	$&$	3200	$&$	1300	$&$	1300	$&$	1700	$&$	3400	$&$	2750	$&$	-2000	$&$	-750	$&$	17,500 	$&$	33	$&$	173.2	$&$	0.0043	$&$	8 \times 10^{-10}	$&$	95\%~{\rm Higgsino}	$	\\	\hline
$	{\rm B }	$&$	1700	$&$	3000	$&$	1300	$&$	1300	$&$	1300	$&$	3700	$&$	2750	$&$	-1900	$&$	1000	$&$	17,500 	$&$	29	$&$	173.7	$&$	0.0060	$&$	9 \times 10^{-10}	$&$	96\%	~{\rm Higgsino}$	\\	\hline
$	{\rm C }	$&$	1300	$&$	4500	$&$	1300	$&$	2500	$&$	1500	$&$	3400	$&$	-750	$&$	-2200	$&$	-2500	$&$	12,500 	$&$	29	$&$	173.7	$&$	0.0015	$&$	2 \times 10^{-10}	$&$	96\%~{\rm Higgsino}	$	\\	\hline
$	{\rm D }	$&$	1500	$&$	3200	$&$	1300	$&$	1700	$&$	1700	$&$	3500	$&$	4500	$&$	-1900	$&$	-750	$&$	12,500 	$&$	33	$&$	174.2	$&$	0.0061	$&$	1.5 \times 10^{-9}	$&$	94\%	~{\rm Higgsino}$	\\	\hline
$	{\rm E }	$&$	1500	$&$	3100	$&$	1300	$&$	1500	$&$	1700	$&$	3500	$&$	-750	$&$	-1900	$&$	-2500	$&$	12,500 	$&$	29	$&$	174.2	$&$	0.0062	$&$	1.5 \times 10^{-9}	$&$	94\%~{\rm Higgsino}	$	\\	\hline
$	{\rm F }	$&$	1500	$&$	3100	$&$	1300	$&$	1500	$&$	1700	$&$	3500	$&$	2750	$&$	-1900	$&$	-2500	$&$	12,500 	$&$	29	$&$	174.2	$&$	0.0045	$&$	1.1 \times 10^{-9}	$&$	94\%	~{\rm Higgsino}$	\\	\hline
$	{\rm G }	$&$	1500	$&$	3200	$&$	1700	$&$	1300	$&$	1700	$&$	3600	$&$	2750	$&$	-2200	$&$	-750	$&$	20,000 	$&$	33	$&$	174.7	$&$	0.0021	$&$	2.6 \times 10^{-10}	$&$	97\%	~{\rm Higgsino}$	\\	\hline
$	{\rm H }	$&$	1500	$&$	3000	$&$	1500	$&$	1700	$&$	1500	$&$	3500	$&$	-2500$&$	-2000	$&$	1000	$&$	20,000 	$&$	33	$&$	174.7	$&$	0.0038	$&$	5.5 \times 10^{-10}	$&$	96\%~{\rm Higgsino}	$	\\	\hline
$	{\rm I }	$&$	1500	$&$	3100	$&$	1500	$&$	1300	$&$	1700	$&$	3500	$&$	-750	$&$	-2000	$&$	-750	$&$	17,500 	$&$	33	$&$	174.7	$&$	0.0064	$&$	1.3 \times 10^{-9}	$&$	95\%	~{\rm Higgsino}$	\\	\hline 
\end{tabular}
\end{table*}

\begin{table*}[htp]
  \centering
  \scriptsize
  \caption{Relevant SUSY spectrum masses for the SUSY breaking soft terms of TABLE~\ref{tab:spectra1} for the \fsu5 D-brane inspired naturalness region. The soft SUSY breaking terms that generate each of these spectra can be identified by the alphabetical label. The light Higgs boson mass $m_h$ column represents the theoretically computed value consisting of the 1-loop and 2-loop SUSY contributions $plus$ the maximum vector-like flippon contribution. The $\Delta M$ columns provide the mass difference between the two SUSY particles given. The $M_{\cal F}$ value is the unification scale where $\alpha_{1X} = {\alpha_5}$. All values in this Table are in GeV.}
\label{tab:spectra2}
\begin{tabular}{|c||c|c|c|c|c|c|c||c|c|c||c|c||c|c|} \hline
$ {\rm Benchmark} $ & $M({\tilde{\chi}^0_1})$ & $M({\tilde{\chi}^0_2})$  &  $M({\tilde{\chi}^\pm_1})$   &  $M({\tilde{t}_1})$  &  $M({\tilde{g}})$ & $M({\tilde{u}_R})$  &  $m_h$  & $\Delta M (\widetilde{t}_1, \widetilde{\chi}_1^0)$ & $\Delta M (\widetilde{\chi}_1^{\pm}, \widetilde{\chi}_1^0)$ & $\Delta M (\widetilde{\chi}_2^0, \widetilde{\chi}_1^0)$ & $ A_t (EW)$ & $ \mu (EW) $ & $ M_{\cal F}$ \\ \hline \hline
$	{\rm A }	$&$	388	$&$	-406	$&$	394	$&$	390	$&$	2011	$&$	3078	$&$	124.01	$&$	1.3	$&$	5.6	$&$	17.6	$&$	3771	$&$	402 	$&$	3.4 \times 10^{17}	$	\\	\hline
$	{\rm B }	$&$	425	$&$	-442	$&$	432	$&$	428	$&$	2254	$&$	3323	$&$	124.64	$&$	2.7	$&$	6.3	$&$	16.4	$&$	4197	$&$	439 	$&$	3.0 \times 10^{17}	$	\\	\hline
$	{\rm C }	$&$	294	$&$	-310	$&$	298	$&$	298	$&$	1732	$&$	2978	$&$	125.40	$&$	4.6	$&$	4.4	$&$	16.4	$&$	3559	$&$	306 	$&$	3.6 \times 10^{17}	$	\\	\hline
$	{\rm D }	$&$	420	$&$	-440	$&$	426	$&$	423	$&$	1997	$&$	3124	$&$	124.53	$&$	3.1	$&$	5.9	$&$	19.7	$&$	3755	$&$	437 	$&$	3.7 \times 10^{17}	$	\\	\hline
$	{\rm E }	$&$	415	$&$	-435	$&$	421	$&$	419	$&$	1996	$&$	3108	$&$	124.96	$&$	3.6	$&$	6.2	$&$	19.9	$&$	3774	$&$	432 	$&$	3.6 \times 10^{17}	$	\\	\hline
$	{\rm F }	$&$	416	$&$	-436	$&$	422	$&$	417	$&$	1996	$&$	3108	$&$	124.87	$&$	0.5	$&$	6.2	$&$	20.0	$&$	3773	$&$	433 	$&$	3.6 \times 10^{17}	$	\\	\hline
$	{\rm G }	$&$	329	$&$	-344	$&$	334	$&$	329	$&$	2017	$&$	3200	$&$	125.26	$&$	0.5	$&$	5.3	$&$	15.3	$&$	3730	$&$	340 	$&$	3.4 \times 10^{17}	$	\\	\hline
$	{\rm H }	$&$	343	$&$	-359	$&$	348	$&$	347	$&$	2016	$&$	3121	$&$	124.47	$&$	4.2	$&$	5.7	$&$	16.1	$&$	3712	$&$	355 	$&$	3.4 \times 10^{17}	$	\\	\hline
$	{\rm I }	$&$	406	$&$	-425	$&$	412	$&$	411	$&$	2012	$&$	3134	$&$	124.46	$&$	4.5	$&$	5.9	$&$	18.7	$&$	3696	$&$	421 	$&$	3.4 \times 10^{17}	$	\\	\hline
\end{tabular}
\end{table*}

\begin{figure*}[htp]
       \centering
        \includegraphics[width=1.0\textwidth]{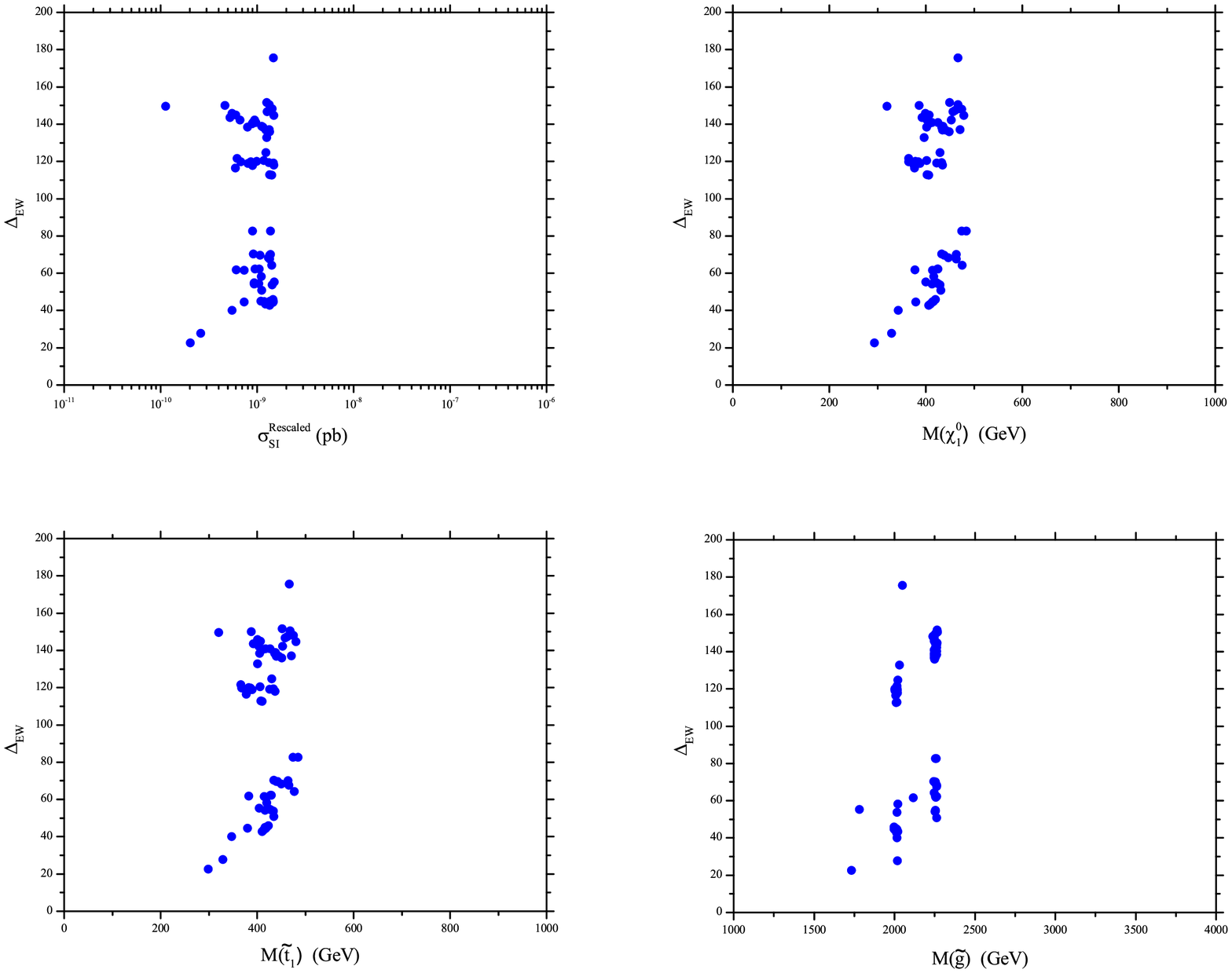}
        \caption{Electroweak fine-tuning measure $\Delta_{EW}$ plot as a function of $\sigma_{SI}^{\rm Rescaled}$, $M(\widetilde{\chi}_1^0)$, $M(\widetilde{t}_1)$, and $M(\widetilde{g})$ for the \fsu5 D-brane inspired naturalness region. The naturalness region in this diagram involves the 74 points that satisfy  $M({\widetilde{g}}) \ge 1.6$~TeV, $124 \le m_h \le 128$~GeV, $\Omega h^2 \le 0.1221$, $\sigma_{SI}^{\rm Rescaled} \le 1.5 \times 10^{-9}$~pb, and $\Delta M(\widetilde{t}_1,\widetilde{\chi}_1^0) \le 5$~GeV. All 74 points depicted here possess an LSP that is at least 92\% Higgsino, but no more than 98\% Higgsino. The $\Delta_{EW}$ stem exclusively from either $M_{H_u}^2 (EW)$ or $\mu^2 (EW)$. These plots highlight a region with $\Delta_{EW} < 30$, regarded as low electroweak fine-tuning.}
        \label{fig:delta_ew}
\end{figure*}

Naturalness demands no disproportionate cancellations amongst the terms within the minimization of the Higgs scalar potential with respect to the $H_u$ and $H_d$ directions. The tree-level minimization condition is

\begin{eqnarray}
\frac{M_Z^2}{2} =
\frac{M_{H_d}^2  -
\tan^2\beta ~M_{H_u}^2}{\tan^2\beta -1} -\mu^2~,~
\label{eq:ewmin}
\end{eqnarray}
however, loop-level radiative corrections to the effective scalar potential $V_{\rm eff} \to V_{\rm tree} + V_{\rm loop}$ deteriorate the situation further as the quadratic $H_u^2$ and $H_d^2$ field coefficients are transformed as $M^2_{H_u} \to M^2_{H_u} + \Sigma_u^u$ and $M^2_{H_d} \to M^2_{H_d} + \Sigma_d^d$. The largest contributions from the radiative corrections $\Sigma_u^u$ and $\Sigma_d^d$ emanate from the top squarks $\widetilde{t}_1$ and $\widetilde{t}_2$, so we will only consider those loop corrections in this study. Provided that we desire no auspicious cancellations on the right-hand side of Eq.~(\ref{eq:ewmin}) in order to produce the correct $Z$-boson mass, we also require a small bilinear Higgs mixing term $\mu$ in addition to the top squarks. Moreover, the quadratic Higgs mass term $M_{H_u}^2$ evolves from a large positive value at the ultimate unification scale $M_{\cal F}$ to a negative value at the EW scale through RGE running due to the large top quark Yukawa coupling, provoking the need for a small negative $M_{H_u}^2$ as well. In summary, the leading culprits to engender contrived results within Eq.~(\ref{eq:ewmin}) are $\widetilde{t}_1$, $\widetilde{t}_2$, $\mu$, and $H_u^2$, motivating minimal values for these quantities. We correspondingly seek regions of the D-brane inspired \fsu5 parameter space yielding small top squarks, small $\mu$ parameter, and a small negative $M_{H_u}^2$ term. A small $\mu$ parameter at the EW scale in turn produces light Higgsinos since the Higgsino mass is near $\mu$, and more practically, a dominant Higgsino component of the LSP. Therefore, we further search for regions of the model space with a dominant Higgsino-like LSP.

The initial step involves a search for an LSP that is greater than 80\% Higgsino. These points are readily recognized by $M(\widetilde{\chi}_2^0) < 0$ on account of the $\mu$ term at $M_{\cal F}$ driven below the gaugino mass terms $M1$ and $M2$ at the electroweak scale via RGE running, sending $\widetilde{\chi}_2^0$ to negative values. Another characteristic of spectra with a Higgsino-like LSP is the compressed nature of the $\widetilde{\chi}_1^0$, $\widetilde{\chi}_1^{\pm}$, and $\widetilde{\chi}_2^0$. The mass deltas expected to produce a Higgsino-like LSP are $\Delta M(\widetilde{\chi}_1^{\pm},\widetilde{\chi}_1^0) \sim 5$~GeV and $\Delta M(\widetilde{\chi}_2^0,\widetilde{\chi}_1^0) \sim 17$~GeV. Accompanying the Higgsino-like LSP, we further require the condition $\Delta M(\widetilde{t}_1,\widetilde{\chi}_1^0) \le 10$~GeV to restrict the results to only those light stops nearly degenerate with the LSP, fulfilling the requisite small light stop limitation. Out of the 400 million points scanned, the intersection of the experimentally viable constraints on $M({\widetilde{g}})$, $m_h$, and $\Omega h^2$ in tandem with an LSP that is $>80\%$ Higgsino and $\Delta M(\widetilde{t}_1,\widetilde{\chi}_1^0) \le 10$~GeV only surrenders $\sim 2900$ points. The resulting region is illustrated in FIG.~\ref{fig:naturalness} and FIG.~\ref{fig:higgsino}, where the dark matter direct detection upper limits on spin-independent neutralino-nucleus cross-sections are superimposed, along with the neutrino scattering floor. All $\sim 2900$ points are discretely depicted in FIG.~\ref{fig:naturalness}, whilst FIG.~\ref{fig:higgsino} delineates smoothly flowing contours of this region highlighting Higgsino percentage of the LSP. It is clear in FIG.~\ref{fig:higgsino} that the more favorable SUSY spectra in terms of smaller spin-independent cross-sections are the larger Higgsino percentages, exhibiting positive accommodation with both characteristics. All points in FIG.~\ref{fig:naturalness} and FIG.~\ref{fig:higgsino} have been rescaled in accordance with Eq. (\ref{eq:omega}).

The analysis from this point forward now enforces two more rather strong restrictions. We want to retain only those points possessing $\sigma_{SI}^{\rm Rescaled} \le 1.5 \times 10^{-9}$~pb, ensuring consistency with the LUX~\cite{Akerib:2016vxi}, PandaX-II~\cite{Tan:2016zwf}, and XENON100~\cite{Aprile:2018dbl} upper limits illuminated in FIG.~\ref{fig:naturalness} and FIG.~\ref{fig:higgsino}. In the region we are exploring here, $\sigma_{SI}^{\rm Rescaled} \sim 1.5 \times 10^{-9}$~pb prevails as an approximate upper limit, so we shall now only consider points less than this boundary. We additionally aim to filter out those points inconsistent with LHC model-independent constraints on $\widetilde{t}_1 \to c \widetilde{\chi}_1^0$. The nearly degenerate light stop and LSP induce a branching fraction of nearly 100\% for $\widetilde{g} \to \widetilde{t}_1 t$ and $\widetilde{t}_1 \to c \widetilde{\chi}_1^0$. However, given the compression between the light stop and LSP, we expect a rather hard top quark but a very soft charm jet, making extraction of this signal from the SM background challenging to say the least. To assist in comparing our naturalness region to the LHC constraints on $\widetilde{t}_1 \to c \widetilde{\chi}_1^0$, post application of $\sigma_{SI}^{\rm Rescaled} \le 1.5 \times 10^{-9}$~pb we overlay the remaining points onto the ATLAS Collaboration exclusion curve plot on $\widetilde{t}_1 \widetilde{t}_1$ production in the monojet search region for the channel $\widetilde{t}_1 \to c \widetilde{\chi}_1^0$, reprinted from Ref.~\cite{Aaboud:2017phn} and displayed in FIG.~\ref{fig:atlas1}. In addition, we superimpose our points onto the ATLAS exclusion curve plot in the charm jets plus zero lepton (0L) search region for the channel $\widetilde{t}_1 \to c \widetilde{\chi}_1^0$, reprinted from Ref.~\cite{Aaboud:2018zjf} and shown in FIG.~\ref{fig:atlas2}. The common element in both these ATLAS Figures is the maximum delta between the light stop and LSP of about 5~GeV, with $\Delta M(\widetilde{t}_1,\widetilde{\chi}_1^0) \lesssim 5$~GeV persisting as viable due to the soft nature of these events and difficulty in differentiation from the SM background. This theme is uniform between both ATLAS and the CMS Collaboration, as the $\widetilde{t}_1 \to c \widetilde{\chi}_1^0$ CMS search regions of Refs.~\cite{Sirunyan:2017kqq,Sirunyan:2017wif,Sirunyan:2018vjp} paint the same picture of viability for $\Delta M(\widetilde{t}_1,\widetilde{\chi}_1^0) \lesssim 5$~GeV. The CMS Ref.~\cite{Sirunyan:2017kiw} for pair production of third-generation squarks states that ``Top squark masses below 510 GeV are excluded for the scenario in which $\widetilde{t}_1 \to c \widetilde{\chi}_1^0$ and the mass splitting between the top squark and the LSP is small'', though Ref.~\cite{Sirunyan:2017kiw} does not explicitly enumerate the value of ``small'', hence we shall consider $\Delta M(\widetilde{t}_1,\widetilde{\chi}_1^0) \lesssim 5$~GeV to remain experimentally viable. The administering of $\sigma_{SI}^{\rm Rescaled} \le 1.5 \times 10^{-9}$~pb and $\Delta M(\widetilde{t}_1,\widetilde{\chi}_1^0) \le 5$~GeV trims the number of residual points from $\sim 2900$ down to only 74 out of 400 million scan! All 74 points have an LSP composition of at least 92\% Higgsino, as FIG.~\ref{fig:higgsino} had indicated, though no point is greater than 98\% Higgsino, supporting small but non-negligible bino and wino components.

We highlight nine benchmark points in TABLES~\ref{tab:spectra1} - \ref{tab:spectra2}. All nine benchmarks are amongst the remaining 74 points satisfying all the constraints applied. It should be noted that the light Higgs boson mass $m_h$ in TABLE~\ref{tab:spectra2} includes all SUSY contributions and the vector-like flippon contribution, lifting the Higgs mass for most of the points to their observed value. This is rather beneficial given the smallness of the light stop and hence its diminished loop-level contribution to the Higgs mass. Notice that there is a repetitive pattern to the $A_t$ and $M^2_{H_u} = M^2_{H_d}$ terms at $M_{\cal F}$  such that we need $A_t \sim -2000$~GeV and $M_{H_u} = M_{H_d}  \sim 3500$~GeV at high scale. This propels consistency within the region for our fine-tuning calculations outlined in the next section.

The entire naturalness model space handily satisfies the B-meson decay and anomalous magnetic moment of the muon boundaries highlighted in the prior section, with our 74 surviving points falling within $3.2 \times 10^{-9} \le Br(B_s^0 \to \mu^+ \mu^-) \le 3.5 \times 10^{-9}$ and $1.5 \times 10^{-10} \le \Delta a_{\mu} \le 2.9 \times 10^{-10}$. However, with regard to the rare b-quark decay, all remaining 74 points compute to $Br(b \to s \gamma) \le 2.34 \times 10^{-4}$, less than the approximate lower $2\sigma$ experimental bound of $Br(b \to s \gamma) \sim 2.77 \times 10^{-4}$, with the smallest of the light stop points returning a value as low as $Br(b \to s \gamma) \sim 10^{-6}$. This is not surprising, given the smallness of the light stop and chargino. The charged heavy Higgs bosons $H^{\pm}$ additionally contribute, but not of sufficient magnitude to offset the minimal SUSY contribution from loops regarding stops and charginos. We emphasize that no points have been excluded from this analysis per the inconsistency with experimental limits on the $Br(b \to s \gamma)$, as we merely note that the SUSY contribution to the total branching ratio is light, thereby suggesting tension with the experimental result.

\section{Fine-tuning}

\begin{figure}[htp]
       \centering
        \includegraphics[width=0.5\textwidth]{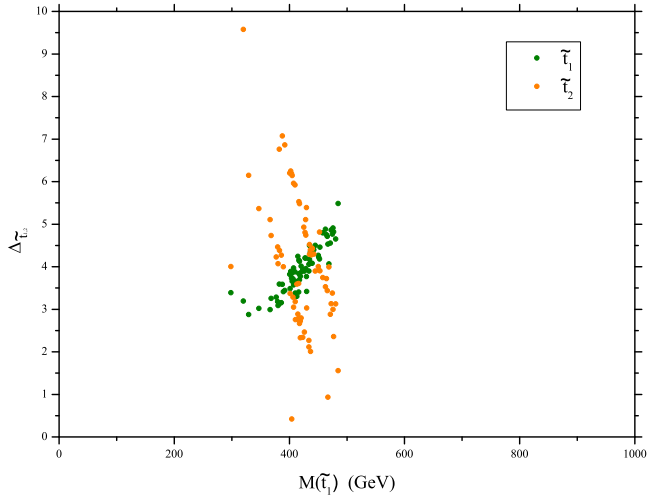}
        \caption{Electroweak fine-tuning measure $\Delta_{\widetilde{t}_{1,2}}$ for only the top squarks $\widetilde{t}_1$ and $\widetilde{t}_2$, plot as a function of $M(\widetilde{t}_1)$ for the \fsu5 D-brane inspired naturalness region. The naturalness region in this diagram involves the 74 points that satisfy  $M({\widetilde{g}}) \ge 1.6$~TeV, $124 \le m_h \le 128$~GeV, $\Omega h^2 \le 0.1221$, $\sigma_{SI}^{\rm Rescaled} \le 1.5 \times 10^{-9}$~pb, and $\Delta M(\widetilde{t}_1,\widetilde{\chi}_1^0) \le 5$~GeV. All 74 points depicted here possess an LSP that is at least 92\% Higgsino, but no more than 98\% Higgsino. This graph shows that the top squarks contribute very little, if any, fine-tuning to the \fsu5 D-brane inspired naturalness region}
        \label{fig:delta_stops}
\end{figure}

It was discussed in the prior section that low fine-tuning conforms with small values for $M(\widetilde{t}_1)$, $M(\widetilde{t}_2)$, $\mu$, and $M_{H_u}^2$, thus we shall conclude this work with an analytical study of how the naturalness region we uncovered here performs in this realm. We follow the prescription offered in Refs.~\cite{Baer:2012up,Baer:2012mv}, calculating measures for electroweak scale fine-tuning. Examining each term on the right-hand side of Eq. (\ref{eq:ewmin}), we have interest in the three electroweak scale tree-level terms 

\begin{eqnarray}
C_{H_u} = \left| \frac{-M^2_{H_u}(EW) {\rm tan}^2 \beta}{{\rm tan}^2 \beta -1} \right|~,~
\label{eq:chu}
\end{eqnarray}

\begin{eqnarray}
C_{H_d} = \left| \frac{M^2_{H_d}(EW)}{{\rm tan}^2 \beta -1} \right|~,~
\label{eq:chd}
\end{eqnarray}

\begin{eqnarray}
C_{\mu} = \left| -\mu^2(EW) \right|~,~
\label{eq:cmu}
\end{eqnarray}
and in the two electroweak scale loop-level terms

\begin{eqnarray}
&C_{\widetilde{t}_{1,2}}&  =  \bigg| \frac{3}{16 \pi^2} M^2_{\widetilde{t}_{1,2}} \left( {\rm log} \frac{M^2_{\widetilde{t}_{1,2}}}{Q^2_{\overline{\widetilde{t}}}} - 1 \right) \nonumber \\
& \times & \left[Y_t^2 - g_Z^2 \mp \frac{Y_t^2 A_t^2 - 8 g_Z^2 (\frac{1}{4} - \frac{2}{3} x_w) \Delta_t}{M^2_{\widetilde{t}_{2}} - M^2_{\widetilde{t}_{1}}} \right] \bigg|~,~
\label{eq:cstops}
\end{eqnarray}
with $\Delta_t = \frac{1}{2}(M^2_{\widetilde{t}_{L}} - M^2_{\widetilde{t}_{R}}) + M_Z^2 {\rm cos} 2 \beta (\frac{1}{4} - \frac{2}{3} x_w)$, $g_Z^2 = \frac{1}{8}(g^2 + g'^2)$, $x_W = {\rm sin}^2 \theta_W$, $Y_t^2 = Y_t^2 (EW)$, and $A_t^2 = A_t^2 (EW)$. For the low-energy scale $Q$, to minimize the logarithms in Eq. (\ref{eq:cstops}) we use the mean of the top squark masses $Q_{\overline{\widetilde{t}}} = (M_{\widetilde{t}_1} + M_{\widetilde{t}_2})/2$. The measure of electroweak scale fine-tuning $\Delta_{EW}$ adopts the maximum of $C_i = \{ C_{H_u}, C_{H_d}, C_{\mu}, C_{\widetilde{t}_{1}}, C_{\widetilde{t}_{2}} \} $, given by

\begin{eqnarray}
\Delta_{EW} =  \frac{{\rm Max} (C_i)}{M_Z^2/2}~.~
\label{eq:cew}
\end{eqnarray}

With our armament of fine-tuning measures in Eqs.~(\ref{eq:chu}) - (\ref{eq:cew}) we now proceed to compute $\Delta_{EW}$ for the 74 points satisfying all criteria outlined in the prior sections. The results of $\Delta_{EW}$ are presented in FIG.~\ref{fig:delta_ew}. We plot them as a function of the primary parameters we are interested in here, namely $\sigma_{SI}^{\rm Rescaled}$, $M(\widetilde{\chi}_1^0)$, $M(\widetilde{t}_1)$, and $M(\widetilde{g})$. The fine-tuning calculations for the benchmark points are featured in TABLE~\ref{tab:spectra3}, inclusive of the percentage of fine-tuning, simply $\Delta^{-1}$. A diminished amount of fine-tuning is preferred since this indicates that all terms on the right-hand side of Eq. (\ref{eq:ewmin}), including radiative corrections, are moving contiguous to the scale of the numerical value of the left-hand side. This is represented by a smaller $\Delta_{EW}$ in TABLE~\ref{tab:spectra3} and FIG.~\ref{fig:delta_ew}. Equivalently, we can also assess success through the percentage of fine-tuning, where a larger percent is preferred, also itemized in TABLE~\ref{tab:spectra3}. Generally speaking, $\Delta_{EW} < 30$, or coequally fine-tuning better than 3\%, is regarded as a low amount of fine-tuning in a SUSY GUT model. The points in FIG.~\ref{fig:delta_ew} present a region with $\Delta_{EW} < 30$, with two benchmarks points in TABLES~\ref{tab:spectra1} - \ref{tab:spectra3} possessing this characteristic. While $\Delta_{EW} < 30$ can be viewed as low fine-tuning, our naturalness region also offers several points with $\Delta_{EW} < 50$. Five of these points are amongst our nine benchmarks points detailed in TABLES~\ref{tab:spectra1} - \ref{tab:spectra3}. The maximum EW term for all 74 points emerges from either $C_{H_u}$ or $C_{\mu}$. 

The electroweak fine-tuning from only the top squarks, as represented by Eq. (\ref{eq:cstops}), is of $\cal{O}$(1), as expected given the small light stop. If we identify the fine-tuning emanating from only the top squarks as

\begin{eqnarray}
\Delta_{\widetilde{t}_{1,2}} =  \frac{ C_{\widetilde{t}_{1,2}}}{M_Z^2/2}~,~
\label{eq:cewstops}
\end{eqnarray}
then we see that all 74 of our points possess $\Delta_{\widetilde{t}_{1,2}} < 10$, indicating that all of the electroweak fine-tuning arises from either $M_{H_u}^2 (EW)$ or $\mu^2 (EW)$. This is illustrated in FIG.~\ref{fig:delta_stops}, showing $\Delta_{\widetilde{t}_{1,2}}$ versus the light stop mass, with the majority of the points having $\Delta_{\widetilde{t}_{1,2}}$ around $\cal{O}$(1). The explicit $\Delta_{\widetilde{t}_{1,2}}$ calculations for the nine benchmark points are provided in TABLE~\ref{tab:spectra3}.

\begin{table}[htp]
  \centering
  \scriptsize
  \caption{Electroweak fine-tuning measures $\Delta_{EW}$ and $\Delta_{\widetilde{t}_{1,2}}$ for the \fsu5 D-brane inspired naturalness region. The fine-tuning percentage is found via $\Delta^{-1}$, also provided here. Those points with $\Delta_{EW} < 30$ are regarded as low electroweak fine-tuning. Note that the fine-tuning for only the top squarks, represented by $\Delta_{\widetilde{t}_{1,2}}$, is of $\cal{O}$(1).}
\label{tab:spectra3}
\begin{tabular}{|c||c|c|c|c|c|c|c|c|c|} \hline
${\rm Benchmark}$&$ {\rm A} $ & $ {\rm B} $ & $ {\rm C} $  &  $ {\rm D} $  &  $ {\rm E} $ &  $ {\rm F} $ &  $ {\rm G} $ &  $ {\rm H} $ &  $ {\rm I} $ \\ \hline \hline
$\Delta_{EW}$&$119$&$ 62 $&$ 23 $&$ 46 $&$ 45 $&$ 45 $&$ 28 $&$ 40 $&$ 43 $\\ \hline
$\Delta_{EW}^{-1}$&$0.8\%$&$ 1.6\% $&$ 4.4\% $&$ 2.2\% $&$ 2.2\% $&$ 2.2\% $&$ 3.6\% $&$ 2.5\% $&$ 2.3\% $\\ \hline \hline
$\Delta_{\widetilde{t}_{1}}$&$3.4$&$ 4.2 $&$ 3.4 $&$ 3.9 $&$ 3.9 $&$ 3.8 $&$ 2.9 $&$ 3.0 $&$ 3.7 $\\ \hline
$\Delta_{\widetilde{t}_{1}}^{-1}$&$29\%$&$ 24\% $&$ 30\% $&$ 25\% $&$ 26\% $&$ 26\% $&$ 35\% $&$ 33\% $&$ 27\% $\\ \hline
$\Delta_{\widetilde{t}_{2}}$&$4.0$&$ 5.1 $&$ 4.0 $&$ 2.3 $&$ 2.7 $&$ 2.8 $&$ 6.1 $&$ 5.4 $&$ 2.8 $\\ \hline
$\Delta_{\widetilde{t}_{2}}^{-1}$&$25\%$&$ 20\% $&$ 25\% $&$ 43\% $&$ 37\% $&$ 36\% $&$ 16\% $&$ 19\% $&$ 36\% $\\ \hline
\end{tabular}
\end{table}

\section{Conclusion}

In the search for SUSY, naturalness has been elevated in significance given its prospects for an elegant natural solution to the hierarchy problems and associated low electroweak fine-tuning. In conjunction, the smallness of the higgsinos and light stops required by naturalness introduces an element of uncertainty into observation of natural models at the LHC given the soft nature of the jets. We examined the well-studied GUT model flipped $SU(5)$ with extra vector-like flippon multiplets, known as \fsu5. However, in this work we allowed freedom on the No-Scale Supergravity boundary conditions at the unification scale, replicating the flipped $SU(5) \times U(1)_X$ GUT representation, referred to as the D-brane inspired model ( ``inspired'' due to its forbidden Yukawa coupling terms).

After a rather comprehensive search for a naturalness sector, we uncovered a region highlighted with points exhibiting a low amount of electroweak fine-tuning, namely $\Delta_{EW} < 30$. The naturalness sector was exposed by constraining the model via $M({\widetilde{g}}) \ge 1.6$~TeV, $124 \le m_h \le 128$~GeV, $\Omega h^2 \le 0.1221$, $\sigma_{SI}^{\rm Rescaled} \le 1.5$~pb, and $\Delta M(\widetilde{t}_1,\widetilde{\chi}_1^0) \le 5$~GeV, providing us with points possessing $\Delta_{EW} < 30$. Attainment of a light Higgs boson mass consistent with the empirically measured value was strengthened by including contributions from the vector-like flippon multiplets, a crucial maneuver given the smallness of the light stops compulsory within naturalness. The resulting region was rather narrow and uniformly supported by nearly 100\% branching fractions for the decay channels $\widetilde{g} \to \widetilde{t}_1 t$ and $\widetilde{t}_1 \to c \widetilde{\chi}_1^0$, indicating the production of a very hard top quark but also a considerably soft charm jet that will be quite difficult to extract from the SM background. Bolstered by these results, we gauged the model against the LHC constraint on $\widetilde{t}_1 \to c \widetilde{\chi}_1^0$, finding that indeed our naturalness region uncovered here does skirt just under the ATLAS and CMS exclusion curves on $\widetilde{t}_1 \to c \widetilde{\chi}_1^0$. 

Could natural SUSY be obscured by the dense Standard Model background in this region heretofore inaccessible at the LHC? Time will tell whether the LHC will yield an affirmative answer to this provocative question. Our imperative here was to merely present a viable physical model that thrives within this elusive space, furnishing motivation to develop enhanced methods of detection for probing concealed SUSY models such as the D-brane inspired model we explored in this work.


\section{Acknowledgments}

Portions of this research were conducted with high performance computational resources provided 
by the Louisiana Optical Network Infrastructure (http://www.loni.org). This research was supported 
in part by the Projects 11475238, 11647601, and 11875062 supported 
by the National Natural Science Foundation of China (TL), 
by the Key Research Program of Frontier Science, Chinese Academy of Sciences (TL),
and by the DOE grant DE-FG02-13ER42020 (DVN). 


\bibliography{bibliography}

\end{document}